\documentstyle[amstex,amssymb]{article}

\catcode`\@=11
\def\numberbysection{\@addtoreset{equation}{section}
        \def\theequation{\thesection.\arabic{equation}}}
\numberbysection
\begin{document}

\newlength{\lno} \lno1.75cm \newlength{\len} \len=\textwidth%
\addtolength{\len}{-\lno}

\setcounter{page}{0}

\baselineskip8mm \renewcommand{\thefootnote}{\fnsymbol{footnote}} \newpage %
\setcounter{page}{0}

\begin{titlepage}     
\vspace{0.5cm}
\begin{center}
{\Large\bf On the ${\cal{U}}_{q}[osp(1|2)]$ Temperley-Lieb model}\\
\vspace{1cm}
{\large \bf  A. Lima-Santos } \\
\vspace{1cm}
{\large \em Universidade Federal de S\~ao Carlos, Departamento de F\'{\i}sica \\
Caixa Postal 676, CEP 13569-905~~S\~ao Carlos, Brasil}\\
\end{center}
\vspace{1.2cm}

\begin{abstract}
This work concerns the boundary integrability of the ${\cal{U}}_{q}[osp(1|2)]$ Temperley-Lieb model. 
We constructed the solutions of the graded reflection equations in order to determine the boundary terms
 of the correspondig spin-1 Hamiltonian. We obtain the eigenvalue expressions as well as its associated 
Bethe ansatz equations by means of the coordinate Bethe ansatz. 
These equations provide the complete description of the spectrum of the model with diagonal integrable 
boundaries.
\end{abstract}
\vspace{2cm}
\begin{center}
\end{center}
\vfill
\begin{center}
\small{\today}
\end{center}
\end{titlepage}

\baselineskip7mm\newpage

\section{Introduction}

Integrability in classical vertex models and quantum spin chains is
intimately connected with solutions of the Yang--Baxter equation \cite%
{BAXTER}. This equation provides an unified approach to construct and study
physical properties of integrable models \cite{FADDEEV, KOREPIN}. Usually
these systems are studied with periodic boundary conditions but more general
boundaries can also be considered.

Although the physical properties associated with the bulk of the system are
not expected to be influenced by boundary conditions in the thermodynamical
limit, there are surface properties such as the interfacial tension where
the boundary conditions are of relevance. Moreover, the conformal spectra of
lattice models at criticality can be modified by the effect of boundaries 
\cite{CARDY}. So, there is an increasing interest in the studies of this
issue.

Integrable systems with open boundary conditions can also be accommodated
within the framework of the Quantum Inverse Scattering Method \cite{SKLYANIN}%
. In addition to the solution of the Yang--Baxter equation governing the
dynamics of the bulk there is another fundamental ingredient, the reflection
matrices \cite{CHEREDNIK}. These matrices, also referred as $K$ matrices,
represent the interactions at the boundaries and compatibility with the bulk
integrability requires these matrices to satisfy the so-called reflection
equations \cite{SKLYANIN, CHEREDNIK}.

The study of general regular solutions of the reflection equations for
vertex models based on $q$-deformed Lie algebras \cite{BAZHANOV, JIMBO} has
been successfully accomplished. See \cite{MALARA} for instance and
references therein. \ Subsequently, similar success has been obtained with
vertex models based on Lie superalgebras \cite{BAZHANOV2, GALLEAS}. For
instance, the\ diagonal solutions associated with the ${\cal U}_{q}[sl(m|n)$ 
\cite{YUE, GE} and ${\cal U}_{q}[osp(2|2)]$ symmetries \cite{GUAN} and
non-diagonal solutions related to super-Yangians $osp(m|n)$ \cite{CRAMPE}
and $sl(m|n)$ \cite{AVAN, GALLEAS2}. The most general set of solutions of
the reflection equation for the vertex models based on Lie superalgebras are
reported in \cite{LIMA}.

More recently, the study of general regular solutions of the reflection
equations for vertex models based on $q$-deformed Temperley-Lieb algebras 
\cite{BATCHELOR} has been done with success \cite{LIMA3, KULISH2, LIMA4}.
The spectra of the corresponding spin chain with integrable open diagonal
boundaries were ontained via the coordinate Bethe ansatz \cite{RIBEIRO}.
However, this same analysis for vertex models based on graded Temperley-Lieb
algebra \cite{ZHANG} is still an open problem.

The aim of this paper is to start to bridge this gap by presenting the most
general set of solutions of the reflection equation for the ${\cal U}%
_{q}[osp(1|2)]$ Temperley-Lieb vertex model and to supply the most general
integrable boundaries for the corresponding spin chain. The description of
the spectrum of the model with diagonal integrable boundaries is obtained
with aid of the Bethe ansatz.

This paper is organized as follows. In the next section we present the $%
{\cal R}$-matrix solution of the ${\cal U}_{q}[osp(1|2)]$ Temperley-Lieb
vertex model. This information allows the way for the analysis of the
corresponding reflection equations and in Section $3$ we present what we
understood to be the most general set of $K$ matrices. In Section $4$ we
present the Temperley-Lieb Hamiltonian with more general integrable boundary
terms, with special interest on diagonal ones. In Section $5$ we use the
generalization of the coordinate Bethe ansatz as presented in \cite{RIBEIRO}%
, in order to obtain directly the spectrum of the Hamiltonian with diagonal
boundary terms. Concluding remarks are discussed in Section $6$.

\section{${\cal R}$ - matrix solution}

\label{basics}

The Temperley-Lieb is a unital associative algebra generated by $%
\{I,U_{1},U_{2},\ldots ,U_{L-1}\}$ subject to the relations%
\begin{eqnarray}
U_{i}^{2} &=&\sqrt{Q}\ U_{i},  \nonumber \\
U_{i}U_{i\pm 1}U_{i} &=&U_{i},  \nonumber \\
U_{i}U_{j} &=&U_{j}U_{i},\qquad |i-j|\ \geq 2  \label{R.1}
\end{eqnarray}%
where $U_{i}$ acts non-trivially in the sites $i$ and $i+1$: 
\begin{equation}
U_{i}=I_{1}\otimes \cdots \otimes I_{i-1}\otimes U\otimes I_{i+2}\otimes
\cdots \otimes I_{L},  \label{R.2}
\end{equation}%
$I$ is the matrix identity and $\sqrt{Q}$ a given number. This tensor
algebra governs the dynamics of a completely integrable model in the sense
that the global quantity%
\begin{equation}
H=\sum_{i=1}^{L-1}U_{i}  \label{R.3}
\end{equation}%
is an involutive integral of motion.

The Temperey-Lieb algebra provided an algebraic framework for constructing
and analyzing different types of integrable lattice models, such as the $Q$%
-state Potts model, {\small IRF} model, $O(n)$ loop model, six vertex model,
etc \cite{MARTIN0}. This equivalence \cite{BAXTER1} was used in order to
obtain the spectra properties of quantum spin chains with periodic boundary
conditions and free ends \cite{KLUMPER}.

Given any representation $U_{i}$, we define the operators%
\begin{equation}
\check{R}_{i}(u)=\frac{\sinh (\eta -u)}{\sinh {\eta }}I_{i}+\frac{\sinh {u}}{%
\sinh {\eta }}U_{i},\qquad i=1,2,...,L-1  \label{R.4}
\end{equation}%
where $\eta $ is related to $Q$ through%
\begin{equation}
2\cosh \eta =\sqrt{Q}  \label{R.5}
\end{equation}%
and it follows from (\ref{R.1}) that%
\begin{eqnarray}
\check{R}_{i}(u)\check{R}_{j}(v) &=&\check{R}_{j}(v)\check{R}_{i}(u),\qquad
|i-j|\ \geq 2  \nonumber \\
\check{R}_{i}(u)\check{R}_{i+1}(u+v)\check{R}_{i}(v) &=&\check{R}_{i+1}(v)%
\check{R}_{i}(u+v)\check{R}_{i+1}(u)  \label{R.6}
\end{eqnarray}%
where $u$ and $v$ are spectral parameters.

From the $Z_{2}$-graded vector space, we refer to $U_{i}$ as a graded vector
representation of the Temperley-Lieb algebra and $\check{R}_{i}(u)$ as the
graded solution of the Yang-Baxter equation (\ref{R.6}).

The orthosympletic ${\cal U}_{q}\left[ osp(M|2n)\right] $ Temperley-Lieb
solutions of the Yang-Baxter equation are well known by Zhang's paper \cite%
{ZHANG}, from which we can write down the ${\cal U}_{q}\left[ osp(1|2)\right]
$ solution: 
\begin{equation}
\check{R}(u)=\frac{\sinh (\eta -u)}{\sinh {\eta }}I+\frac{\sinh {u}}{\sinh {%
\eta }}U,  \label{R.7}
\end{equation}%
where $I$ is the $9$ by $9$ matrix identity and the Temperley-Lieb operator 
\begin{equation}
U=\left( 
\begin{array}{ccc|ccc|ccc}
0 & 0 & 0 & 0 & 0 & 0 & 0 & 0 & 0 \\ 
0 & 0 & 0 & 0 & 0 & 0 & 0 & 0 & 0 \\ 
0 & 0 & -q^{-1} & 0 & q^{-\frac{1}{2}} & 0 & 1 & 0 & 0 \\ \hline
0 & 0 & 0 & 0 & 0 & 0 & 0 & 0 & 0 \\ 
0 & 0 & -q^{-\frac{1}{2}} & 0 & 1 & 0 & q^{\frac{1}{2}} & 0 & 0 \\ 
0 & 0 & 0 & 0 & 0 & 0 & 0 & 0 & 0 \\ \hline
0 & 0 & 1 & 0 & -q^{\frac{1}{2}} & 0 & -q & 0 & 0 \\ 
0 & 0 & 0 & 0 & 0 & 0 & 0 & 0 & 0 \\ 
0 & 0 & 0 & 0 & 0 & 0 & 0 & 0 & 0%
\end{array}%
\right) ,  \label{R.8}
\end{equation}%
which, in a spin chain language, is the projector onto the two-sites spin
zero singlet written in the spin-$1$ basis $\{\left\vert +\right\rangle
,\left\vert 0\right\rangle ,\left\vert -\right\rangle \}$. Here we have used
the grading {\small FBF {\it i.e.}},{\small \ \ }$[\left\vert +\right\rangle
]]=[\left\vert -\right\rangle ]=1$ and $[\left\vert 0\right\rangle ]=0$.

\bigskip Using the relation ${\cal R}=P_{{\rm g}}\check{R}$ where $P_{{\rm g}%
}$ is the graded permutation operator%
\begin{equation}
P_{{\rm g}}\left\vert \alpha \right\rangle \otimes \left\vert \beta
\right\rangle =(-1)^{[\alpha ][\beta ]}\left\vert \beta \right\rangle
\otimes \left\vert \alpha \right\rangle  \label{R.9}
\end{equation}%
with $\alpha ,\beta =+,0,-$, the graded ${\cal R-}$ matrix has the form%
\begin{equation}
{\cal R}=\left( 
\begin{array}{ccccccccc}
-x_{1} & 0 & 0 & 0 & 0 & 0 & 0 & 0 & 0 \\ 
0 & 0 & 0 & x_{1} & 0 & 0 & 0 & 0 & 0 \\ 
0 & 0 & -x_{2} & 0 & \sqrt{q}x_{2} & 0 & qx_{2}-x_{1} & 0 & 0 \\ 
0 & x_{1} & 0 & 0 & 0 & 0 & 0 & 0 & 0 \\ 
0 & 0 & -\frac{1}{\sqrt{q}}x_{2} & 0 & x_{1}+x_{2} & 0 & \sqrt{q}x_{2} & 0 & 
0 \\ 
0 & 0 & 0 & 0 & 0 & 0 & 0 & x_{1} & 0 \\ 
0 & 0 & -x_{1}+\frac{1}{q}x_{2} & 0 & -\frac{1}{\sqrt{q}}x_{2} & 0 & -x_{2}
& 0 & 0 \\ 
0 & 0 & 0 & 0 & 0 & x_{1} & 0 & 0 & 0 \\ 
0 & 0 & 0 & 0 & 0 & 0 & 0 & 0 & -x_{1}%
\end{array}%
\right)  \label{R.10}
\end{equation}%
where%
\begin{equation}
x_{1}=\frac{\sinh (\eta -u)}{\sinh \eta }\quad {\rm and}\quad x_{2}=\frac{%
\sinh (u)}{\sinh \eta },  \label{R.11}
\end{equation}%
satisfying a regular condition ${\cal R}(0)=P_{{\rm g}}$.

This ${\cal R}$ - matrix defines the local structure of Boltzmann weight of\
a graded $15$-vertex model in a two-dimensional lattice. It follows that we
can define a row-to-row transfer matrix $t(u)$ as super-trace of a monodromy
matrix ${\cal T}_{{\cal A}}(u)$%
\begin{eqnarray}
t(u) &=&{\rm str}_{{\cal A}}({\cal T}_{{\cal A}}(u)),  \nonumber \\
{\cal T}_{{\cal A}}(u) &=&{\cal R}_{L{\cal A}}(u){\cal R}_{L-1{\cal A}%
}(u)\cdots {\cal R}_{2{\cal A}}(u){\cal R}_{1{\cal A}}(u),  \label{R.12}
\end{eqnarray}%
Here the notation means that the operator ${\cal R}_{i{\cal A}}(u)$ is a
matrix in the auxiliary space ${\cal A}$ corresponding to the horizontal
degrees of freedom and its matrix elements are operators on the quantum
space $\otimes _{i=1}^{L}V_{i}$ , where $V_{i}$ represents the vertical
space of states and $i$ the site of one-dimensional lattice of size $L$.

The transfer matrix (\ref{R.12}) is the functional generator of infinite
conserved quantities%
\begin{equation}
t(u)=\exp \left( \sum_{k=1}^{\infty }Q_{k}u^{k-1}\right)  \label{R.13}
\end{equation}%
The commutation relation $[t(u),t(v)]=0$ , $u\neq v$ \ is provided by (\ref%
{R.6}) and it follows that $[Q_{k},Q_{l}]=0$, $\forall k,l$. \ Here we note
that $Q_{2}$ can be identified with the global Hamiltonian (\ref{R.3}),
which define the integrable closed spin chain.

The spectra of the orthosympletic Temperley-Lieb Hamiltonian with Martin's
boundary condition \cite{MARTIN0}, periodic boundary condition and free ends
are known \cite{LIMA0}.

\section{K - matrix solution}

The notion of quantum integrability was extended to work with open boundary
problems \cite{SKLYANIN}. In addition to the graded ${\cal R}$-matrix
describes the bulk dynamics, we have to introduce reflection $K$ matrices to
describe such boundary conditions. These new matrices represent the
interactions at the right and left ends of the open spin chain. This is a
consequence of the reflection equation, which reads 
\begin{equation}
{\cal R}_{12}(u-\mu )K_{1}^{(-)}(u){\cal R}_{21}(u+\mu )K_{2}^{(-)}(\mu
)=K_{2}^{(-)}(\mu ){\cal R}_{12}(u+\mu )K_{1}^{(-)}(u){\cal R}_{21}(u-\mu ).
\label{K.1}
\end{equation}

In the case of open boundary conditions, the graded transfer matrix can be
written as the super-trace 
\begin{equation}
t(u)={\rm str}_{{\cal A}}\left[ K_{{\cal A}}^{(+)}(u){\cal T}_{{\cal A}%
}(u)K_{{\cal A}}^{(-)}(u)\left[ {\cal T}_{{\cal A}}(-u)\right] ^{-1}\right] ,
\label{K.2}
\end{equation}%
where $K_{{\cal A}}^{(-)}(u)$ can be chosen as one of the solutions of the
reflection equation (\ref{K.1}). The other boundary matrix $K_{{\cal A}%
}^{(+)}(u)$ is obtained from the previous one by means of the isomorphism 
\cite{NEPO}, 
\begin{equation}
K_{{\cal A}}^{(+)}(u)=K_{{\cal A}}^{(-)}(-u-\rho )^{{\rm st}}M,  \label{K.3}
\end{equation}%
where $\rho =-\eta $ is the crossing parameter and ${\rm st}$ means
super-transposition.

For the ${\cal U}_{q}\left[ osp(1|2)\right] $ Temperley-Lieb model the
graded $M$ matrix is given by 
\begin{equation}
M=\left( 
\begin{array}{ccc}
q^{-1} & 0 & 0 \\ 
0 & 1 & 0 \\ 
0 & 0 & q%
\end{array}%
\right) .  \label{K.4}
\end{equation}

The integrable open spin chain is obtained by means of the logarithmic
derivative of the transfer matrix (\ref{K.2}), such that, 
\begin{equation}
H=\sum_{k=1}^{L-1}U_{k,k+1}+\frac{\sinh {\eta }}{2}\frac{dK_{1}^{(-)}(u)}{du}%
\Big|_{u=0}+\frac{{\rm str}_{{\cal A}}{\left[ K_{{\cal A}}^{(+)}(0)U_{L{\cal %
A}}\right] }}{{\rm str}_{{\cal A}}{\left[ K_{{\cal A}}^{(+)}(0)\right] }}.
\label{K.5}
\end{equation}%
Now, we begin to solve the reflection equation (\ref{K.1}) for the ${\cal U}%
_{q}\left[ osp(1|2)\right] $ Temperley-Lieb vertex model in order to obtain
the boundary terms of (\ref{K.5}).

Using (\ref{R.10}), the reflection matrix $\left( K^{(-)}(u)\right)
_{ij}=k_{ij}(u),\ i,j=\{1,2,3\}$, with $K_{1}^{(-)}(u)=K^{(-)}(u)\otimes I$
, $K_{2}^{(-)}(u)=I\otimes K^{(-)}(u)$, ${\cal R}_{12}={\cal R}$ and ${\cal R%
}_{21}=P_{{\rm g}}{\cal R}P_{{\rm g}}$, the matrix equation (\ref{K.1}) has $%
81$ functional relations for the $k_{ij}(u)$ matrix elements, many of them
not independent relations. In order to solve these functional equations, we
shall proceed as follows. First we consider the $(i,j)$ component of the
matrix equation (\ref{K.1}). By differentiating it with respect to $v$ and
taking $v=0$, we get algebraic equations involving the single variable $u$
and nine parameters 
\begin{equation}
\beta _{ij}=\frac{dk_{ij}(u)}{du}\Big|_{u=0},\quad i,j=1,2,3.  \label{K.7}
\end{equation}

Analyzing the reflection equations one can see that they possess a special
structure. Several equations exist involving only two non-diagonal elements.
They can be solved by the relations%
\begin{eqnarray}
k_{12}(u) &=&\beta _{12}\frac{k_{13}(u)}{\beta _{13}},  \nonumber \\
k_{21}(u) &=&\beta _{21}\frac{k_{13}(u)}{\beta _{13}},\quad k_{23}(u)=\beta
_{23}\frac{k_{13}(u)}{\beta _{13}},  \nonumber \\
k_{31}(u) &=&\beta _{31}\frac{k_{13}(u)}{\beta _{13}},\quad k_{32}(u)=\beta
_{32}\frac{k_{13}(u)}{\beta _{13}}.  \label{K.8}
\end{eqnarray}%
We are thus left with several equations involving two diagonal elements and $%
k_{13}(u)$. From the equations $(1,2)$ and $(1,4)$ we have%
\begin{eqnarray}
k_{22}(u) &=&k_{11}(u)+\left( \beta _{22}-\beta _{11}\right) \frac{k_{13}(u)%
}{\beta _{13}},  \nonumber \\
k_{33}(u) &=&k_{11}(u)+\left( \beta _{33}-\beta _{11}\right) \frac{k_{13}(u)%
}{\beta _{13}},  \label{K.9}
\end{eqnarray}%
respectively.

Finally, we can use the equation $(1,3)$ to find $k_{11}(u)$:%
\begin{eqnarray}
k_{11}(u) &=&\frac{k_{13}(u)}{\beta _{13}\left( x_{2}(u)\cosh \eta
+x_{1}(u)\right) }\left\{ \frac{x_{1}(u)x_{2}^{\prime }(u)-x_{1}^{\prime
}(u)x_{2}(u)}{x_{2}(u)}\right.  \nonumber \\
&&\left. -\frac{\beta _{12}\beta _{23}}{2\beta _{13}}x_{1}(u)-\frac{\beta
_{22}-\beta _{11}}{2}x_{2}(u)-\frac{\beta _{33}-\beta _{11}}{2}\left(
x_{1}(u)-qx_{2}(u)\right) \right\}  \nonumber \\
&&  \label{K.10}
\end{eqnarray}%
where $x_{i}^{\prime }(u)=dx_{i}(u)/du,$ $i=1,2$.

Now, substituting these expressions into the remaining equations $(i,j)$, we
are left with several constraint equations involving the $\beta _{ij}$
parameters.

From the equation $(2,3)$ we can choose%
\begin{equation}
\beta _{22}-\beta _{11}=\frac{\beta _{12}\beta _{23}}{\beta _{13}}-\frac{%
\beta _{21}\beta _{13}}{\beta _{23}},  \label{K.11}
\end{equation}%
and from the equation $(3,7)$ 
\begin{equation}
\beta _{33}-\beta _{11}=\frac{\beta _{13}\beta _{32}}{\beta _{12}}-\frac{%
\beta _{21}\beta _{13}}{\beta _{23}}.  \label{K.12}
\end{equation}%
Here we note that $\beta _{11}$is fixed by the normal condition, $%
K^{(-)}(0)=I$. \ Moreover, all the remaining constraint equations are solved
by the symmetric relation%
\begin{equation}
\beta _{32}\beta _{21}\beta _{13}=\beta _{23}\beta _{12}\beta _{31}.
\label{K.13}
\end{equation}%
from which we can fix $\beta _{32}$ in function of $\beta _{12},\beta
_{13},\beta _{21},\beta _{23}$ and $\beta _{31}$. In this way we have
obtained a five free parameter solution of (\ref{K.1}) for\ the ${\cal U}_{q}%
\left[ osp(1|2)\right] $ Temperley-Lieb vertex model.

The corresponding $K^{(+)}(u)$ reflection matrix is obtained by the
isomorphism (\ref{K.3}), namely%
\begin{equation}
K^{(+)}(u)=\left( 
\begin{array}{ccc}
\frac{1}{q}k_{11}(-u+\eta ) & k_{21}(-u+\eta ) & qk_{31}(-u+\eta ) \\ 
-\frac{1}{q}k_{12}(-u+\eta ) & k_{22}(-u+\eta ) & -qk_{32}(-u+\eta ) \\ 
\frac{1}{q}k_{13}(-u+\eta ) & k_{23}(-u+\eta ) & qk_{33}(-u+\eta )%
\end{array}%
\right)  \label{K.14}
\end{equation}%
where we have used the super-transpostion $(A^{st})_{\alpha \beta
}=(-1)^{[\alpha ][\beta ]+[\beta ]}A_{\beta \alpha }$ and the graded $M$
matrix (\ref{K.4}) but, replacing all parameters $\beta _{ij}$ by the new
parameters $\alpha _{ij}$.

From the general solution (\ref{K.8}) \ to (\ref{K.13}) one can see $%
k_{13}(u)$ as an arbitrary function satisfying the normal condition.
Therefore the choice%
\begin{eqnarray}
k_{13}(u) &=&\beta _{13}x_{2}(u)\frac{x_{2}(u)\cosh \eta +x_{1}(u)}{%
x_{1}(u)x_{2}^{\prime }(u)-x_{1}^{\prime }(u)x_{2}(u)}  \nonumber \\
&=&\frac{1}{2}\beta _{13}\sinh (2u)  \label{K.15a}
\end{eqnarray}%
does not imply any restriction as compared to the general case, but simplify
our $K^{-}(u)$ - matrix to%
\[
K^{-}(u)= 
\]%
\begin{equation}
\left( 
\begin{array}{ccc}
k_{11} & \frac{1}{2}\beta _{12}\sinh (2u) & \frac{1}{2}\beta _{13}\sinh (2u)
\\ 
\frac{1}{2}\beta _{21}\sinh (2u) & k_{11}+\frac{1}{2}(\beta _{22}-\beta
_{11})\sinh (2u) & \frac{1}{2}\beta _{23}\sinh (2u) \\ 
\frac{1}{2}\beta _{31}\sinh (2u) & \frac{1}{2}\beta _{32}\sinh (2u) & k_{11}+%
\frac{1}{2}(\beta _{33}-\beta _{11})\sinh (2u)%
\end{array}%
\right)  \label{K.16a}
\end{equation}%
and%
\begin{eqnarray}
k_{11} &=&k_{11}(u,\beta )=1-\frac{1}{2}\left\{ \frac{\beta _{12}\beta _{23}%
}{2\beta _{13}}x_{1}(u)+\frac{1}{2}(\beta _{22}-\beta _{11})x_{2}(u)\right. 
\nonumber \\
&&+\left. \frac{1}{2}(\beta _{33}-\beta _{11})\left(
x_{1}(u)-qx_{2}(u)\right) \right\} x_{2}(u)\sinh \eta  \label{K.17}
\end{eqnarray}%
where $\beta _{22}-\beta _{11}$, $\beta _{33}-\beta _{11}$ and $\beta _{32}$
are given by (\ref{K.11}), (\ref{K.12}) and (\ref{K.13}), respectively.

\subsection{Reduced $K$- matrix solutions}

For particular choice of the free parameters in (\ref{K.16a}) to (\ref{K.17}%
), we can derive several reduced solutions. For instance, making $\beta
_{21}=\beta _{31}=\beta _{32}=0$, the $K^{(-)}$ - matrix (\ref{K.16a}) is
reduced to a three free parameters solution, $K_{{\rm up}}^{(-)}$ - matrix,
the {\rm up}-triangular solution. Similarly, making $\beta _{12}=\beta
_{13}=\beta _{23}=0$ we get a three free parameters $K_{{\rm down}}^{(-)}$ -
matrix, the {\rm down}-triangular solution. The corresponding $K^{(+)}$
matrices are obtained from (\ref{K.14}). However, in order to obtain all
diagonal solutions, it is simpler to solve the reflection equations again.

Taking into account diagonal $K$ matrices, all reflection equations (\ref%
{K.1}) are solved when we find $k_{22}(u)$ and $k_{33}(u)$ as functions of $%
k_{11}(u)$ provided that the diagonal parameters $\beta _{ii}$ satisfy the
constraint equation%
\begin{equation}
(\beta _{33}-\beta _{22})(\beta _{33}-\beta _{11})(\beta _{22}-\beta _{11})=0
\label{K.15}
\end{equation}%
From (\ref{K.15}), of the three matrix elements, two have the same value.
Let us normalize one of them to be equal to $1$ such that the other entry is
given by%
\begin{equation}
k_{pp}(u)=\frac{\beta _{pp}\ x_{2}(u)[\Delta
_{1}x_{2}(u)+x_{1}(u)]+2[x_{1}(u)x_{2}^{\prime }(u)-x_{1}^{\prime
}(u)x_{2}(u)]}{\beta _{pp}\ x_{2}(u)[\Delta
_{2}x_{2}(u)+x_{1}(u)]-2[x_{1}(u)x_{2}^{\prime }(u)-x_{1}^{\prime
}(u)x_{2}(u)]}  \label{K.16}
\end{equation}%
where $\Delta _{1}+\Delta _{2}=-q^{-1}+1-q$.

Identifying the diagonal indexes by $(1,2,3)\circeq (q^{-1},1,q)$ one can
see that $\Delta _{1}$ is the sum of terms corresponding to the positions of
the entries $1$ and $\Delta _{2}$ is equal to sum of terms corresponding to
the positions of the entries $k_{pp}(u)$. It means that we have six $%
K^{(-)}(u)$ diagonal solutions, namely%
\begin{equation}
D_{1}^{[I]}={\rm diag}\left( k_{11},1,1\right) ,\ D_{2}^{[I]}={\rm diag}%
\left( 1,k_{22},1\right) ,\ D_{3}^{[I]}={\rm diag}\left( 1,1,k_{22}\right)
\label{K.17a}
\end{equation}%
with two entries equal to $1$ and%
\begin{eqnarray}
k_{11}(u) &=&\frac{\beta _{11}\
x_{2}(u)[(1-q)x_{2}(u)+x_{1}(u)]+2[x_{1}(u)x_{2}^{\prime }(u)-x_{1}^{\prime
}(u)x_{2}(u)]}{\beta _{11}\
x_{2}(u)[-q^{-1}x_{2}(u)+x_{1}(u)]-2[x_{1}(u)x_{2}^{\prime
}(u)-x_{1}^{\prime }(u)x_{2}(u)]},  \nonumber \\
k_{22}(u) &=&\frac{\beta _{22}\
x_{2}(u)[(-q^{-1}-q)x_{2}(u)+x_{1}(u)]+2[x_{1}(u)x_{2}^{\prime
}(u)-x_{1}^{\prime }(u)x_{2}(u)]}{\beta _{22}\
x_{2}(u)[x_{2}(u)+x_{1}(u)]-2[x_{1}(u)x_{2}^{\prime }(u)-x_{1}^{\prime
}(u)x_{2}(u)]},  \nonumber \\
k_{33}(u) &=&\frac{\beta _{33}\
x_{2}(u)[(1-q^{-1})x_{2}(u)+x_{1}(u)]+2[x_{1}(u)x_{2}^{\prime
}(u)-x_{1}^{\prime }(u)x_{2}(u)]}{\beta _{33}\
x_{2}(u)[-qx_{2}(u)+x_{1}(u)]-2[x_{1}(u)x_{2}^{\prime }(u)-x_{1}^{\prime
}(u)x_{2}(u)]}.  \label{K.18}
\end{eqnarray}%
The second type has only one entry equal to $1$%
\begin{equation}
D_{1}^{[II]}={\rm diag}\left( 1,k_{22},k_{22}\right) ,\ D_{2}^{[II]}={\rm %
diag}\left( k_{33},1,k_{33}\right) ,\ D_{3}^{[II]}={\rm diag}\left(
k_{11},k_{11},1\right)  \label{K.19}
\end{equation}%
where%
\begin{eqnarray}
k_{11}(u) &=&\frac{\beta _{11}\
x_{2}(u)[-qx_{2}(u)+x_{1}(u)]+2[x_{1}(u)x_{2}^{\prime }(u)-x_{1}^{\prime
}(u)x_{2}(u)]}{\beta _{11}\
x_{2}(u)[(1-q^{-1})x_{2}(u)+x_{1}(u)]-2[x_{1}(u)x_{2}^{\prime
}(u)-x_{1}^{\prime }(u)x_{2}(u)]},  \nonumber \\
k_{22}(u) &=&\frac{\beta _{22}\
x_{2}(u)[-q^{-1}x_{2}(u)+x_{1}(u)]+2[x_{1}(u)x_{2}^{\prime
}(u)-x_{1}^{\prime }(u)x_{2}(u)]}{\beta _{22}\
x_{2}(u)[(1-q)x_{2}(u)+x_{1}(u)]-2[x_{1}(u)x_{2}^{\prime }(u)-x_{1}^{\prime
}(u)x_{2}(u)]},  \nonumber \\
k_{33}(u) &=&\frac{\beta _{33}\
x_{2}(u)[x_{2}(u)+x_{1}(u)]+2[x_{1}(u)x_{2}^{\prime }(u)-x_{1}^{\prime
}(u)x_{2}(u)]}{\beta _{33}\
x_{2}(u)[(-q^{-1}-q)x_{2}(u)+x_{1}(u)]-2[x_{1}(u)x_{2}^{\prime
}(u)-x_{1}^{\prime }(u)x_{2}(u)]}.  \label{K.20}
\end{eqnarray}%
Again, the prime means derivative in respect to $u$ and $x_{i}(u),$ $i=1,2$
are given by (\ref{R.11}). Note also that the difference between these one
free parameter solutions comes from the partitions of $2\cosh \eta
=-q^{-1}+1-q$. Moreover, we have the symmetry $q\leftrightarrow q^{-1}$ and
the corresponding six $K^{(+)}(u)$ diagonal solutions are obtained by the
isomorphism (\ref{K.3}).

\section{${\cal U}_{q}[osp(1|2)]$ Temperley-Lieb spin chain}

The boundary terms of Hamiltonian (\ref{K.5}) are directly obtained from the 
$K^{(\pm )}$ matrices. In particular, the boundary term acting non-trivially
in the site $1$ has the form 
\begin{eqnarray}
bt_{1} &=&\frac{\sinh \eta }{2}\frac{dK^{(-)}(u)}{du}\Big|_{u=0}  \nonumber
\\
&=&\frac{1}{2}\sinh \eta \left( 
\begin{array}{ccc}
t_{11} & \beta _{12} & \beta _{13} \\ 
\beta _{21} & t_{22} & \beta _{23} \\ 
\beta _{31} & \beta _{32} & t_{33}%
\end{array}%
\right)  \label{spin.1}
\end{eqnarray}%
where%
\begin{eqnarray}
t_{11} &=&\frac{1}{2}\left( -\frac{\beta _{12}\beta _{23}}{\beta _{13}}-%
\frac{\beta _{23}\beta _{31}}{\beta _{21}}+\frac{\beta _{21}\beta _{13}}{%
\beta _{23}}\right)  \nonumber \\
t_{22} &=&\frac{1}{2}\left( \frac{\beta _{12}\beta _{23}}{\beta _{13}}-\frac{%
\beta _{23}\beta _{31}}{\beta _{21}}-\frac{\beta _{21}\beta _{13}}{\beta
_{23}}\right)  \nonumber \\
t_{33} &=&\frac{1}{2}\left( -\frac{\beta _{12}\beta _{23}}{\beta _{13}}+%
\frac{\beta _{23}\beta _{31}}{\beta _{21}}-\frac{\beta _{21}\beta _{13}}{%
\beta _{23}}\right)  \label{spin.2}
\end{eqnarray}%
and we remember that $\beta _{32}$ is given by (\ref{K.13}).

\bigskip The boundary term acting non-trivially in the site $L$ is given by%
\begin{eqnarray}
bt_{L} &=&\frac{{\rm str}_{{\cal A}}[K_{{\cal A}}^{(+)}(0)U_{L{\cal A}}]}{%
{\rm str}_{{\cal A}}[K_{{\cal A}}^{(+)}(0)]}  \nonumber \\
&=&\frac{1}{2}\sinh (\eta )\left( 
\begin{array}{ccc}
\frac{v_{11}}{\sinh (2\eta )} & q^{\frac{1}{2}}\alpha _{32} & -q\alpha _{31}
\\ 
-q^{-\frac{1}{2}}\alpha _{23} & \frac{v_{22}}{\sinh (2\eta )} & q^{\frac{1}{2%
}}\alpha _{21} \\ 
-q^{-1}\alpha _{13} & -q^{-\frac{1}{2}}\alpha _{12} & \frac{v_{33}}{\sinh
(2\eta )}%
\end{array}%
\right)  \label{spin.3}
\end{eqnarray}%
where 
\begin{eqnarray}
v_{11} &=&2+\left( -\frac{\alpha _{12}\alpha _{23}}{\alpha _{13}}+(1-q^{-1})%
\frac{\alpha _{23}\alpha _{31}}{\alpha _{21}}+q^{-1}\frac{\alpha _{21}\alpha
_{13}}{\alpha _{23}}\right) \sinh (\eta )  \nonumber \\
v_{22} &=&2+\left( -(q+q^{-1})\frac{\alpha _{12}\alpha _{23}}{\alpha _{13}}+q%
\frac{\alpha _{23}\alpha _{31}}{\alpha _{21}}+q^{-1}\frac{\alpha _{21}\alpha
_{13}}{\alpha _{23}}\right) \sinh \eta  \nonumber \\
v_{33} &=&2+\left( -\frac{\alpha _{12}\alpha _{23}}{\alpha _{13}}+q\frac{%
\alpha _{23}\alpha _{31}}{\alpha _{21}}+(1-q)\frac{\alpha _{21}\alpha _{13}}{%
\alpha _{23}}\right) \sinh \eta  \label{spin.4}
\end{eqnarray}

Here we have new five free parameters $\alpha _{12},\alpha _{13},\alpha
_{21},\alpha _{23}$ and $\alpha _{31}$. Note that $\alpha _{32}$ is given by
(\ref{K.13}), replacing $\beta $ by $\alpha $.

With these expressions we have obtained the general integrable boundaries
terms for the quantum spin-$1$ chain associated with \ the ${\cal U}_{q}%
\left[ osp(1|2)\right] $ Temperley-Lieb vertex model.

\subsection{Diagonal boundaries}

The choice $\beta _{21}=\beta _{31}=\beta _{32}=0$ ($\alpha _{21}=\alpha
_{31}=\alpha _{32}=0$) in $t_{b1}$ ($t_{bL}$) gives us the {\rm up}%
-triangular right boundary and {\rm down}-triangular left boundary both with
three free parameters and vice-versa for the choice $\beta _{12}=\beta
_{13}=\beta _{23}=0$ ($\alpha _{12}=\alpha _{13}=\alpha _{23}=0$).

In the next section we will use the coordinate Bethe ansatz in order to
diagonalize the Hamiltonian (\ref{K.5}) with diagonal integrable boundaries.
Therefore, let us write explicitly its diagonal entries.

From right $K^{(-)}$-matrices $D_{1}^{[I]},D_{2}^{[I]}$ and $D_{3}^{[I]}$,
we can compute the type $I$ boundaries 
\begin{eqnarray}
bt_{1}^{(1,[I])} &=&{\rm diag}(r(\beta ),0,0),\quad bt_{L}^{(1,[I])}={\rm %
diag}(s_{1}(\alpha ),s_{1}(\alpha ),t_{1}(\alpha ))  \nonumber \\
bt_{1}^{(2,[I])} &=&{\rm diag}(0,r(\beta ),0),\quad bt_{L}^{(2,[I])}={\rm %
diag}(s_{2}(\alpha ),t_{2}(\alpha ),s_{2}(\alpha ))  \nonumber \\
bt_{1}^{(3,[I])} &=&{\rm diag}(0,0,r(\beta )),\quad bt_{L}^{(3,[I])}={\rm %
diag}(t_{3}(\alpha ),s_{3}(\alpha ),s_{3}(\alpha ))  \label{diag.1}
\end{eqnarray}%
where $r(\beta )=\frac{1}{2}\beta \sinh \eta $ and $\beta $ is the free
parameter. \ For each $bt_{L}^{(i,[I])}$ we have $s_{i}(\alpha )$ and $%
t_{i}(\alpha ),i=1,2,3$ given by%
\begin{eqnarray}
s_{1}(\alpha ) &=&\frac{2+q^{-1}\alpha \sinh \eta }{4\cosh \eta },\quad
s_{2}(\alpha )=\frac{2-\alpha \sinh \eta }{4\cosh \eta },  \nonumber \\
s_{3}(\alpha ) &=&\frac{2+q\alpha \sinh \eta }{4\cosh \eta }  \label{diag.2}
\end{eqnarray}%
and 
\begin{eqnarray}
t_{1}(\alpha ) &=&\frac{2+(1-q)\alpha \sinh \eta }{4\cosh \eta },\quad
t_{2}(\alpha )=\frac{2-(q+q^{-1})\alpha \sinh \eta }{4\cosh \eta }, 
\nonumber \\
t_{3}(\alpha ) &=&\frac{2+(1-q^{-1})\alpha \sinh \eta }{4\cosh \eta }
\label{diag.3}
\end{eqnarray}%
where $\alpha $ is the corresponding free parameter.

We have more three solutions corresponding to the $K^{(-)}$ matrices (\ref%
{K.19})%
\begin{eqnarray}
bt_{1}^{(1,[II])} &=&{\rm diag}(0,r(\beta ),r(\beta )),\quad
bt_{L}^{(1,[II])}={\rm diag}(s_{1}(-\alpha ),s_{1}(-\alpha ),t_{1}(-\alpha
)),  \nonumber \\
bt_{1}^{(2,[II])} &=&{\rm diag}(r(\beta ),0,r(\beta )),\quad
bt_{L}^{(2,[II])}={\rm diag}(s_{2}(-\alpha ),t_{2}(-\alpha ),s_{2}(-\alpha
)),  \nonumber \\
bt_{1}^{(3,[II])} &=&{\rm diag}(r(\beta ),r(\beta ),0),\quad
bt_{L}^{(3,[II])}={\rm diag}(t_{3}(-\alpha ),s_{3}(-\alpha ),s_{3}(-\alpha
)).  \label{diag.4}
\end{eqnarray}

It follows that we have $6$ different integrable boundaries related by the
isomorphism (\ref{K.3}). However, it is worth to note that other combination
of the boundaries are allowed $%
B_{1,L}^{(i,j,[a,b])}=bt_{1}^{(j,[b])}+bt_{L}^{(i,[a])}$ with $i,j=1,2,3$, $%
a,b=I,II$ resulting in $36$ integrable boundaries for the spin-$1$ ${\cal U}%
_{q}[osp(1|2)]$ Temperley-Lieb Hamiltonian.

The action of the boundary terms on the Hilbert space is given by 
\[
B_{1,L}^{(i,j,[a,b])}\left\vert \overset{1}{\sigma }\cdots \overset{L}{\tau }%
\right\rangle ={\cal E}_{\sigma \tau }^{(i,j,[a,b])}|\overset{1}{\sigma }%
\cdots \overset{L}{\tau }> 
\]%
where ${\cal E}_{\sigma \tau }^{(i,j,[a,b])}=l_{\sigma \sigma
}^{(i,[a])}+r_{\tau \tau }^{(j,[b])}$ and the sites are indexed by $\sigma
,\tau =(1,2,3)\doteq (+,0,-)$. Here we recall that $l_{\sigma \sigma
}^{(i,[a])}$ and $r_{\tau \tau }^{(j,[b])}$ are the matrix elements of $%
bt_{1}^{(i,[a])}$ and $bt_{L}^{(j,[b])}$ respectively.

In the next section, we will restrict ourselves to the case of integrable
boundaries related by the isomorphism ($%
B_{1,L}^{(i,[a])}=bt_{1}^{(i,[a])}+bt_{L}^{(i,[a])}$) and we shall use the
coordinate Bethe ansatz in order to obtain the eigenvalues of the
Hamiltonian (\ref{K.5}).

\section{Coordinate Bethe ansatz}

\label{cBA}

We known that the bulk part of the Hamiltonian (\ref{K.5}) is the projector
operator onto the two-site spin zero. This implies that there exist $3\times
2^{L-1}$ states in a lattice with $L$ sites which are eigenstates of the
bulk Hamiltonian with zero eigenvalues. However, these states are also
eigenstates of the boundary part of the Hamiltonian $B_{1,L}^{(i,[a])}$ with
eigenvalues ${\cal E}_{\sigma \tau }^{(i,[a])}$. For instance, in a lattice
with $L=4$ sites we have the $24$ natural reference states (Grouping
according to its ${\cal E}_{ab}$):%
\begin{equation}
\left. 
\begin{array}{c}
\left\vert +++\right\rangle \\ 
\left\vert +0++\right\rangle \\ 
\left\vert ++0+\right\rangle%
\end{array}%
\right\} :{\cal E}_{11},\quad \left. 
\begin{array}{c}
\left\vert 0+++\right\rangle \\ 
\left\vert 0+0+\right\rangle \\ 
\left\vert 0-0+\right\rangle%
\end{array}%
\right\} :{\cal E}_{21},\quad \left. 
\begin{array}{c}
\left\vert -0++\right\rangle \\ 
\left\vert --++\right\rangle%
\end{array}%
\right\} :{\cal E}_{31},  \label{cba.1a}
\end{equation}%
\begin{equation}
\left. 
\begin{array}{c}
\left\vert +++0\right\rangle \\ 
\left\vert +0+0\right\rangle \\ 
\left\vert +0-0\right\rangle%
\end{array}%
\right\} :{\cal E}_{12},\quad \left. 
\begin{array}{c}
\left\vert 0++0\right\rangle \\ 
\left\vert 0-+0\right\rangle%
\end{array}%
\right\} :{\cal E}_{22},\quad \left. 
\begin{array}{c}
\left\vert ---0\right\rangle \\ 
\left\vert -0-0\right\rangle \\ 
\left\vert -0+0\right\rangle%
\end{array}%
\right\} :{\cal E}_{32},  \label{cba.1b}
\end{equation}%
\begin{equation}
\quad \left. 
\begin{array}{c}
\left\vert ++0-\right\rangle \\ 
\left\vert +0--\right\rangle%
\end{array}%
\right\} :{\cal E}_{13},\quad \left. 
\begin{array}{c}
\left\vert 0---\right\rangle \\ 
\left\vert 0-0-\right\rangle \\ 
\left\vert 0+0-\right\rangle%
\end{array}%
\right\} :{\cal E}_{23},\quad \left. 
\begin{array}{c}
\left\vert ---\right\rangle \\ 
\left\vert --0-\right\rangle \\ 
\left\vert -0--\right\rangle%
\end{array}%
\right\} :{\cal E}_{33}.  \label{cba.1c}
\end{equation}

Moreover, apart from the natural degenerescence of the boundary eigenvalues $%
{\cal E}_{\sigma \tau }^{(i,[a])}$, one can see from the structure of the
boundary matrix $K^{(\pm )}$ that not all ${\cal E}_{\sigma \tau }^{(i,[a])}$
are independent. More precisely, with aid of (\ref{diag.1}) and (\ref{diag.4}%
) we have 
\begin{eqnarray}
{\cal E}_{31}^{(1,[a])} &=&\left\{ 
\begin{array}{c}
t_{1}(\alpha )+r(\beta ) \\ 
t_{1}(-\alpha )%
\end{array}%
\right. ,  \nonumber \\
{\cal E}_{21}^{(1,[a])} &=&{\cal E}_{11}^{(1,[a])}=\left\{ 
\begin{array}{c}
s_{1}(\alpha )+r(\beta ) \\ 
s_{1}(-\alpha )%
\end{array}%
\right. ,  \nonumber \\
{\cal E}_{32}^{(1,[a])} &=&{\cal E}_{33}^{(1,[a])}=\left\{ 
\begin{array}{c}
t_{1}(\alpha ) \\ 
t_{1}(-\alpha )+r(\beta )%
\end{array}%
\right. ,  \nonumber \\
{\cal E}_{12}^{(1,[a])} &=&{\cal E}_{22}^{(1,[a])}={\cal E}_{13}^{(1,[a])}=%
{\cal E}_{23}^{(1,[a])}=\left\{ 
\begin{array}{c}
s_{1}(\alpha ) \\ 
s_{1}(-\alpha )+r(\beta )%
\end{array}%
\right. ,  \label{cba.2}
\end{eqnarray}%
for the first solutions ($1^{{\rm st}}$), 
\begin{eqnarray}
{\cal E}_{22}^{(2,[a])} &=&\left\{ 
\begin{array}{c}
t_{2}(\alpha )+r(\beta ) \\ 
t_{2}(-\alpha )%
\end{array}%
\right. ,  \nonumber \\
{\cal E}_{12}^{(2,[[a])} &=&{\cal E}_{32}^{(2,[[a])}=\left\{ 
\begin{array}{c}
s_{2}(\alpha )+r(\beta ) \\ 
s_{2}(-\alpha )%
\end{array}%
\right. ,  \nonumber \\
{\cal E}_{21}^{(2,[[a])} &=&{\cal E}_{23}^{(2,[[a])}=\left\{ 
\begin{array}{c}
t_{2}(\alpha ) \\ 
t_{2}(-\alpha )+r(\beta )%
\end{array}%
\right. ,  \nonumber \\
{\cal E}_{13}^{(2,[[a])} &=&{\cal E}_{11}^{(2,[[a])}={\cal E}%
_{31}^{(2,[[a])}={\cal E}_{33}^{(2,[[a])}=\left\{ 
\begin{array}{c}
s_{2}(\alpha ) \\ 
s_{2}(-\alpha )+r(\beta )%
\end{array}%
\right. ,  \label{cba.3}
\end{eqnarray}%
for the second solutions ($2^{{\rm nd}}$) \ and 
\begin{eqnarray}
{\cal E}_{13}^{(3,[a])} &=&\left\{ 
\begin{array}{c}
t_{3}(\alpha )+r(\beta ) \\ 
t_{3}(-\alpha )%
\end{array}%
\right. ,  \nonumber \\
{\cal E}_{12}^{(3,[[a])} &=&{\cal E}_{11}^{(3,[[a])}=\left\{ 
\begin{array}{c}
s_{3}(\alpha )+r(\beta ) \\ 
s_{3}(-\alpha )%
\end{array}%
\right. ,  \nonumber \\
{\cal E}_{23}^{(3,[[a])} &=&{\cal E}_{33}^{(3,[[a])}=\left\{ 
\begin{array}{c}
t_{3}(\alpha ) \\ 
t_{3}(-\alpha )+r(\beta )%
\end{array}%
\right. ,  \nonumber \\
{\cal E}_{21}^{(3,[[a])} &=&{\cal E}_{22}^{(3,[[a])}={\cal E}%
_{31}^{(3,[[a])}={\cal E}_{32}^{(3,[[a])}=\left\{ 
\begin{array}{c}
s_{3}(\alpha ) \\ 
s_{3}(-\alpha )+r(\beta )%
\end{array}%
\right. ,  \label{cba.4}
\end{eqnarray}%
for the third solutions ($3^{{\rm th}}$).

Note that the up expressions in each ($\left\{ {}\right. $ ) of (\ref{cba.2}%
), (\ref{cba.3}) and (\ref{cba.4}) correspond to type $I$ solution while the
down expressions correspond to type $II$ solution.

In face of the large number of reference states, the standard construction
of the all eigenstates seems to be impracticable. However, in order to
obtain the eigenvalues of the Hamiltonian it is enough to work out with a
few reference states. In fact, we can take one reference state from each
block of these eigenvalues ${\cal E}_{\sigma \tau }^{(i,[a])}$ \cite{RIBEIRO}%
. From now on, we drop the label for different solutions of the reflection
equation from the boundary eigenvalues, such that ${\cal E}_{\sigma \tau
}^{(i,[a])}={\cal E}_{\sigma \tau }$.

\subsection{Ferromagnetic reference state}

We shall start by considering the pseudo particle as a singlet over any of
the reference state listed above and extended to $L$ sites ($\left\vert
++\cdots ++\right\rangle $: ${\cal E}_{11}$). In general, it is convenient
to start our ansatz with the following linear combination of the basis
states \cite{LIMA2}, 
\begin{eqnarray}
\left\vert \Omega (k)\right\rangle &=&\sum_{i=-1}^{1}\epsilon (i)q^{-\frac{%
i+1}{2}}\left\vert k(i,-i)\right\rangle  \nonumber \\
&\equiv &-\frac{1}{q}\left\vert \cdots \overset{k}{{\bf +}}{\bf -}\cdots
\right\rangle +\frac{1}{\sqrt{q}}\left\vert \cdots \overset{k}{0}0\cdots
\right\rangle +\left\vert \cdots \overset{k}{{\bf -}}{\bf +}\cdots
\right\rangle  \label{cba.5}
\end{eqnarray}%
where $(\cdots )$ means that the remained sites are defined by the reference
state considered. Here $\epsilon (-1)=\epsilon (0)=1$, $\epsilon (1)=-1$ and 
$1\leq k\leq L-1$. It follows that $\left\vert \Omega (k)\right\rangle $\ is
an eigenstate of $U_{k,k+1}$ such that 
\begin{align}
& U_{k,k+1}\left\vert \Omega (k)\right\rangle =\sqrt{Q}\left\vert \Omega
(k)\right\rangle ,\qquad U_{k,k+1}\left\vert \Omega (k\pm 1)\right\rangle
=\left\vert \Omega (k)\right\rangle ,  \nonumber \\
& U_{k,k+1}\left\vert \Omega (j)\right\rangle =0,~\mbox{if }k\neq
\{j-1,j,j+1\},  \label{cba.6}
\end{align}%
where $\sqrt{Q}=-q^{-1}+1-q$.

The action of the Hamiltonian $H=\sum_{k=1}^{L-1}U_{k,k+1}+B_{1,L}$ over
this state results 
\begin{eqnarray}
H\left\vert \Omega (k)\right\rangle &=&\left( \sqrt{Q}+{\cal E}_{11}\right)
\left\vert \Omega (k)\right\rangle +\left\vert \Omega (k-1)\right\rangle
+\left\vert \Omega (k+1)\right\rangle ,  \nonumber \\
1 &<&k<L-1.  \label{cba.7}
\end{eqnarray}%
where $B_{1,L}\left\vert \Omega (k)\right\rangle ={\cal E}_{11}\left\vert
\Omega (k)\right\rangle $ for $1<k<L-1$, due to ferromagnetic reference
state.

In addition, we have for $k=1$ 
\begin{equation}
H\left\vert \Omega (1)\right\rangle =\left( \sqrt{Q}+{\cal E}_{11}\right)
\left\vert \Omega (1)\right\rangle +\left\vert \Omega (2)\right\rangle
+(B_{1,L}-{\cal E}_{11})\left\vert \Omega (1)\right\rangle ,  \label{cba.8}
\end{equation}%
and for $k=L-1$ 
\begin{equation}
H\left\vert \Omega (L-1)\right\rangle =\left( \sqrt{Q}+{\cal E}_{11}\right)
\left\vert \Omega (L-1)\right\rangle +\left\vert \Omega (L-2)\right\rangle
+(B_{1,L}-{\cal E}_{11})\left\vert \Omega (L-1)\right\rangle ,  \label{cba.9}
\end{equation}%
The equations (\ref{cba.8}) and (\ref{cba.9}) can be seen as extensions of (%
\ref{cba.7}) provided we define two new states%
\begin{eqnarray}
\left\vert \Omega (0)\right\rangle &=&(B_{1,L}-{\cal E}_{11})\left\vert
\Omega (1)\right\rangle  \nonumber \\
&=&q^{-\frac{1}{2}}\left( {\cal E}_{21}-{\cal E}_{11}\right) \left\vert
00+\cdots +\right\rangle +\left( {\cal E}_{31}-{\cal E}_{11}\right)
\left\vert -++\cdots +\right\rangle  \label{cba.10}
\end{eqnarray}%
and%
\begin{eqnarray}
\left\vert \Omega (L)\right\rangle &=&(B_{L,1}-{\cal E}_{11})\left\vert
\Omega (L-1)\right\rangle  \nonumber \\
&=&-q^{-1}\left( {\cal E}_{13}-{\cal E}_{11}\right) \left\vert +\cdots
+-\right\rangle +q^{-\frac{1}{2}}\left( {\cal E}_{13}-{\cal E}_{11}\right)
\left\vert +\cdots +00\right\rangle  \label{cba.11}
\end{eqnarray}%
From the action of $H$ on these new states we have two closing relations

\begin{eqnarray}
H\left\vert \Omega (0)\right\rangle &=&\Delta _{l}^{(1)}\left\vert \Omega
(1)\right\rangle +B_{1,L}\left\vert \Omega (0)\right\rangle  \nonumber \\
\Delta _{l}^{(1)} &=&({\cal E}_{21}-{\cal E}_{11})-q({\cal E}_{31}-{\cal E}%
_{11})  \label{cba.12}
\end{eqnarray}%
and 
\begin{eqnarray}
H\left\vert \Omega (L)\right\rangle &=&\Delta _{r}^{(1)}\left\vert \Omega
(L-1)\right\rangle +B_{1,L}\left\vert \Omega (L)\right\rangle ,  \nonumber \\
\Delta _{r}^{(1)} &=&({\cal E}_{12}-{\cal E}_{11})-q^{-1}({\cal E}_{13}-%
{\cal E}_{11})  \label{cba.13}
\end{eqnarray}

It follows from (\ref{cba.10}) that the action of $B_{1,L}$ over $\left\vert
\Omega (0)\right\rangle $ depend on the possible choices of the boundary
eigenvalues 
\begin{eqnarray}
B_{1,L}\left\vert \Omega (0)\right\rangle &=&\left\{ 
\begin{array}{c}
{\cal E}_{31}\left\vert \Omega (0)\right\rangle \quad {\rm if}\quad {\cal E}%
_{21}={\cal E}_{11}\quad (1^{{\rm st}}) \\ 
{\cal E}_{21}\left\vert \Omega (0)\right\rangle \quad {\rm if}\quad {\cal E}%
_{31}={\cal E}_{11}\quad (2^{{\rm nd}}) \\ 
{\cal E}_{21}\left\vert \Omega (0)\right\rangle \quad {\rm if}\quad {\cal E}%
_{31}={\cal E}_{21}\quad (3^{{\rm th}})%
\end{array}%
\right.  \nonumber \\
&\doteq &{\cal E}_{(v_{1})_{i}1}\left\vert \Omega (0)\right\rangle ,\quad
i=1,2,3  \label{cba.14}
\end{eqnarray}%
where we have used the vector notation $v_{1}=(3,2,2)$ whose components $%
i\in \{1,2,3\}$ represent the different solutions of the reflection
equations $(\ref{cba.2},\ref{cba.3},\ref{cba.4})$, in this order.

Similarly, the eigenvalue problem $B_{1,L}\left\vert \Omega (L)\right\rangle 
$ depend on the possible choices of the boundary eigenvalues in (\ref{cba.11}%
)%
\begin{eqnarray}
B_{1,L}\left\vert \Omega (L)\right\rangle &=&\left\{ 
\begin{array}{c}
{\cal E}_{13}\left\vert \Omega (L)\right\rangle \quad {\rm if}\quad {\cal E}%
_{12}={\cal E}_{11}\quad (3^{{\rm th}}) \\ 
{\cal E}_{12}\left\vert \Omega (L)\right\rangle \quad {\rm if}\quad {\cal E}%
_{13}={\cal E}_{11}\quad (2^{{\rm nd}}) \\ 
{\cal E}_{12}\left\vert \Omega (L)\right\rangle \quad {\rm if}\quad {\cal E}%
_{12}={\cal E}_{13}\quad (1^{{\rm st}})%
\end{array}%
\right.  \nonumber \\
&\doteq &{\cal E}_{1(u_{1})_{j}}\left\vert \Omega (L)\right\rangle ,\quad
j=1,2,3  \label{cba.15}
\end{eqnarray}%
where $u_{1}=(2,2,3)$.

Taking into account these relations, valid for the ferromagnetic reference
state, we can reconstruct all steps of the coordinate Bethe ansatz presented
in the reference \cite{RIBEIRO}.

\subsection{One-particle state}

In the first non-trivial sector, we assume the following ansatz for the
eigenstates 
\begin{equation}
\Psi _{1}=\sum_{k=1}^{L-1}A(k)\left\vert \Omega (k)\right\rangle .
\label{one.1}
\end{equation}

Imposing the eigenvalue equation $H\Psi _{1}=E_{1}\Psi _{1}$ is fulfilled,
we obtain a set of equations for the function $A(k)$%
\begin{equation}
\left( E_{1}-\sqrt{Q}-{\cal E}_{11}\right) A(k)=A(k-1)+A(k+1),\quad 1<k<L-1
\label{one.2}
\end{equation}%
and its extensions to include $k=1$ and $k=L-1$%
\begin{equation}
(E_{1}-{\cal E}_{(v_{1})_{i}1})A(0)=\Delta _{l}^{(1)}A(1)  \label{one.3}
\end{equation}%
and%
\begin{equation}
(E_{1}-{\cal E}_{1(u_{1})_{i}})A(L)=\Delta _{l}^{(1)}A(L-1)  \label{one.4}
\end{equation}

Taking the ansatz for the plane wave amplitude 
\begin{equation}
A(k)=a(\theta )\xi ^{k}-a(-\theta )\xi ^{-k},  \label{one.5}
\end{equation}%
we have the following expression for the energy eigenvalues 
\begin{equation}
E_{1}(\xi )={\cal E}_{11}+\sqrt{Q}+\xi +\xi ^{-1}.  \label{one.6}
\end{equation}%
After we fix the parameter $\xi =e^{{\rm i}\theta }$ and the ratio of the
amplitudes $a(\theta )/a(-\theta )$ we are left with the Bethe ansatz
equation 
\begin{eqnarray}
\xi ^{2L} &=&\left( \frac{E_{1}(\theta )-{\cal E}_{(v_{1})_{i}1}-\xi \Delta
_{l}^{(1)}}{E_{1}(\theta )-{\cal E}_{(v_{1})_{i}1}-\xi ^{-1}\Delta _{l}^{(1)}%
}\right) \left( \frac{E_{1}(\theta )-{\cal E}_{1(u_{1})_{i}}-\xi \Delta
_{r}^{(1)}}{E_{1}(\theta )-{\cal E}_{1(u_{1})_{i}}-\xi ^{-1}\Delta _{r}^{(1)}%
}\right)  \nonumber \\
&\equiv &F_{l}(\theta )F_{r}(\theta )  \label{one.7}
\end{eqnarray}%
where $E_{1}(\theta )=\sqrt{Q}+{\cal E}_{11}+2\cos \theta $.

\subsection{Two-particle state}

In the next particle sector, we have two interacting pseudo-particles, which
can be represented as a product of two pseudo-particles eigenstates, as
given by 
\begin{equation}
\Psi _{2}=\sum_{k_{1}+1<k_{2}}A(k_{1},k_{2})\left\vert \Omega
(k_{1},k_{2})\right\rangle ,  \label{two.1}
\end{equation}%
It follows from the notation for one-particle state that 
\begin{equation}
\left\vert \Omega (k_{1},k_{2})\right\rangle =\sum_{i,j=-1}^{1}\epsilon
(i)\epsilon (j)q^{-\frac{i+j+2}{2}}\left\vert
k_{1}(i,-i);k_{2}(j,-j)\right\rangle .  \label{two.2}
\end{equation}%
where $\epsilon (1)=-1$ and $\epsilon (0)=\epsilon (-1)=1$.

We can split the action of the Hamiltonian on the state $\left\vert \Omega
(k_{1},k_{2})\right\rangle $ in four cases: (i) The case where two
pseudo-particles are separated in the bulk, (ii) The case where the
pseudo-particles are separated but one of them or both are at the boundaries
(iii) The case where the particles are neighbours in the bulk (iv) The case
where the particles are neighbours at the boundaries. Similarly to
one-particle state we need to introduce new states \ and see the action of
the Hamiltonian on them. These equation are not reported here, but we can
follow the construction of two-particle states of \cite{RIBEIRO} in order to
verify the results reported below.

Take into account the eigenvalue equation ($H\Psi _{2}=E_{2}\Psi _{2}$), one
can obtain the two-particle eigenvalue 
\begin{equation}
E_{2}=2\sqrt{Q}+{\cal E}_{11}+\xi _{1}+\xi _{1}^{-1}+\xi _{2}+\xi _{2}^{-1},
\end{equation}%
provided that the following parametrization for the plane wave amplitudes is
assumed 
\begin{equation}
A(k_{1},k_{2})=\sum_{P}\varepsilon _{P}a(\theta _{1},\theta _{2})\xi
_{1}^{k_{1}}\xi _{2}^{k_{2}},
\end{equation}%
where the sum extends over all permutations and negations of momenta ($%
\theta _{i}$), such that $\xi _{i}=e^{{\rm i}\theta _{i}}$, and $\varepsilon
_{P}$ is the signature of permutations and negations. This structure already
reflects the existence of the boundary reflections.

From the phase shift relations we obtain the corresponding Bethe ansatz
equations%
\begin{equation}
\xi _{1}^{2L}=F_{l}(\theta _{1})F_{r}(\theta _{1})\left( \frac{s(\theta
_{1},\theta _{2})}{s(\theta _{2},\theta _{1})}\right) \left( \frac{s(\theta
_{2},-\theta _{1})}{s(-\theta _{1},\theta _{2})}\right)
\end{equation}%
and%
\begin{equation}
\xi _{2}^{2L}=F_{l}(\theta _{2})F_{r}(\theta _{2})\left( \frac{s(\theta
_{2},\theta _{1})}{s(\theta _{1},\theta _{2})}\right) \left( \frac{s(\theta
_{1},-\theta _{2})}{s(-\theta _{2},\theta _{1})}\right) .
\end{equation}%
The defining relations for the boundary factors are%
\begin{equation}
F_{l}(\theta _{a})=\left( \frac{E_{1}(\theta _{a})-{\cal E}%
_{(v_{1})_{i}1}-\xi _{a}\Delta _{l}^{(1)}}{E_{1}(\theta _{a})-{\cal E}%
_{(v_{1})_{i}1}-\xi _{a}^{-1}\Delta _{l}^{(1)})}\right)
\end{equation}%
and%
\begin{equation}
F_{r}(\theta _{a})=\left( \frac{E_{1}(\theta _{a})-{\cal E}%
_{1(u_{1})_{i}}-\xi _{a}\Delta _{r}^{(1)}}{E_{1}(\theta _{a})-{\cal E}%
_{1(u_{1})_{i}}-\xi _{a}^{-1}\Delta _{r}^{(1)})}\right)
\end{equation}%
where $a=1,2$. Moreover%
\begin{equation}
E_{1}(\theta _{a})=\sqrt{Q}+{\cal E}_{11}+2\cos \theta _{a}
\end{equation}%
and%
\begin{equation}
s(\theta _{a},\theta _{b})=1+\xi _{a}\xi _{b}+\xi _{a}\sqrt{Q},\quad a\neq b
\end{equation}

\subsection{$m$-particle state}

The generalization to any number $m$ of pseudo-particles goes along the same
lines as before. Therefore, we just present the final results.

The eigenstates are obtained as a product of $m$ pseudo-particle eigenstates
(\ref{cba.5}) 
\begin{equation}
\Psi _{m}=\sum_{\{k_{i}+1<k_{i+1}\}}A(k_{1},\cdots ,k_{m})\left\vert \Omega
(k_{1},\ldots ,k_{m})\right\rangle
\end{equation}%
where%
\begin{eqnarray}
&&\left\vert \Omega (k_{1},\ldots ,k_{m})\right\rangle  \nonumber \\
&=&\sum_{\{i_{1},\cdots ,i_{m}\}=-1}^{1}\epsilon (i_{i})\cdots \epsilon
(i_{m})q^{-\frac{i_{1}+\cdots i_{m}+m}{2}}\left\vert
k_{1}(i_{1},-i_{1}),\ldots ,k_{m}(i_{m},-i_{m})\right\rangle
\end{eqnarray}%
with the signs $\epsilon (-1)=\epsilon (0)=1$ and $\epsilon (1)=-1$. The
energy eigenvalues are given by the sum of single pseudo-particle energies 
\begin{equation}
E_{m}={\cal E}_{11}+\sum_{a=1}^{m}\left( \sqrt{Q}+\xi _{a}+\xi
_{a}^{-1}\right) ,
\end{equation}%
where $m$ ranges from $0$ to $L/2$, and the corresponding Bethe ansatz
equations depend on the phase shift of two pseudo-particles and on the
boundary factors: 
\begin{eqnarray}
\xi _{a}^{2L} &=&F_{l}(\theta _{a})F_{r}(\theta _{a})\prod_{\overset{b=1}{%
b\neq a}}^{m}\left( \frac{s(\theta _{a},\theta _{b})}{s(\theta _{b},\theta
_{a})}\right) \left( \frac{s(\theta _{b},-\theta _{a})}{s(-\theta
_{a},\theta _{b})}\right)  \nonumber \\
a &=&1,2,...,m.
\end{eqnarray}

\subsection{Other reference states}

We already known from \cite{RIBEIRO} that to obtain the whole spectrum of
the Hamiltonian we have to consider additional reference states. This has to
be done for each different boundary eigenvalues ${\cal E}_{\sigma \tau }$.
As a result of that, we must have as many as reference states and
consequently Bethe ansatz equations as boundary eigenvalues.

In principle, we have nine boundary eigenvalues ${\cal E}_{\sigma \tau }$.
If one choose one reference state for each boundary eigenvalues (e.g the
first state of each block of \ref{cba.1a} extended to $L$-sites) and proceed
along the same lines as the previous subsection, we obtain nine eigenvalues
expressions 
\begin{equation}
E_{m}^{(\sigma ,\tau )}={\cal E}_{\sigma \tau }+\sum_{a=1}^{m}\sqrt{Q}+\xi
_{a}+\xi _{a}^{-1},
\end{equation}%
as well as its associated Bethe ansatz equations 
\begin{equation}
\xi _{a}^{2L}=F_{l}^{(\sigma ,\tau )}(\theta _{a})F_{r}^{(\sigma ,\tau
)}(\theta _{a})\prod_{\overset{b=1}{b\neq a}}^{m}\left( \frac{s(\theta
_{a},\theta _{b})}{s(\theta _{b},\theta _{a})}\right) \left( \frac{s(\theta
_{b},-\theta _{a})}{s(-\theta _{a},\theta _{b})}\right) ,
\end{equation}%
with 
\begin{equation}
F_{l}^{(\sigma ,\tau )}(\theta _{a})=\left( \frac{E_{1}^{(\sigma ,\tau
)}(\theta _{a})-{\cal E}_{(v_{\sigma })_{i}\tau }-\xi _{a}\Delta
_{l}^{(\sigma )}}{E_{1}^{(\sigma ,\tau )}(\theta _{a})-{\cal E}_{(v_{\sigma
})_{i}\tau }-\xi _{a}^{-1}\Delta _{l}^{(\sigma )}}\right) ,
\end{equation}%
and%
\begin{equation}
F_{r}^{(\sigma ,\tau )}(\theta _{a})=\left( \frac{E_{1}^{(\sigma ,\tau
)}(\theta _{a})-{\cal E}_{\sigma (u_{\tau })_{i}}-\xi _{a}\Delta _{r}^{(\tau
)}}{E_{1}^{(\sigma ,\tau )}(\theta _{a})-{\cal E}_{\sigma (u_{\tau
})_{i}}-\xi _{a}^{-1}\Delta _{r}^{(\tau )}}\right)
\end{equation}%
where 
\begin{equation}
E_{1}^{(\sigma ,\tau )}(\theta _{a})=\sqrt{Q}+{\cal E}_{\sigma \tau }+2\cos
\theta _{a}
\end{equation}%
The $\Delta _{\{l,r\}}^{(\sigma )}$ means that we are considering the
reference states listed in (\ref{cba.1a}), extended to $L$-sites \ and
identified by ${\cal E}_{\sigma 1}$. Explictly, the reference state $%
\left\vert ++\cdots ++\right\rangle $: ${\cal E}_{11}$ (ferromagnetic), the
reference state $\left\vert 0+\cdots ++\right\rangle $: ${\cal E}_{21}$%
\begin{equation}
\Delta _{l}^{(2)}=-({\cal E}_{31}-{\cal E}_{21})q-({\cal E}_{11}-{\cal E}%
_{21})q^{-1},\quad \Delta _{r}^{(2)}=-({\cal E}_{11}-{\cal E}_{12})q-({\cal E%
}_{13}-{\cal E}_{12})q^{-1}
\end{equation}%
and the reference state $\left\vert -+\cdots ++\right\rangle $: ${\cal E}%
_{31}$%
\begin{equation}
\Delta _{l}^{(3)}=({\cal E}_{21}-{\cal E}_{31})-({\cal E}_{11}-{\cal E}%
_{31})q^{-1},\quad \Delta _{r}^{(3)}=({\cal E}_{12}-{\cal E}_{13})-({\cal E}%
_{11}-{\cal E}_{13})q.
\end{equation}%
The remaining index $(v_{\sigma })_{i}$ are defined by $v_{2}=(3,1,1)$, $%
v_{3}=(1,2,1)$ and the $(u_{\tau })_{i}$ are given by $u_{2}=(1,1,3)$, $%
u_{3}=(1,2,1)$.

It is not all because we also have to consider the reamined reference states
(\ref{cba.1b}) and (\ref{cba.1c}). However, we can see from (\ref{cba.2}) to
(\ref{cba.4}) that most of these equations degenerate into each other,
resulting in four equations for each integrable boundary.

\section{Conclusion}

\label{CONCLUSION}

In this paper the reflection $K$ - matrix solution of the spin-$1$ ${\cal U}%
_{q}\left[ osp(1|2)\right] $ Temperley-Lieb vertex model is presented. This
step paves the way for the analysis of the corresponding open Hamiltonian
for which we present what we hope to be the most general set of integrable
boundary terms. We obtained the spectrum of the spin-$1$ ${\cal U}_{q}\left[
osp(1|2)\right] $ Temperley-Lieb spin chain with diagonal open boundary
conditions. We have identified that this model has large number of possible
reference states. By selecting a small subset of these states, we manage to
obtain four eigenvalue expressions and its associated Bethe ansatz equations
by means of a generalization of the coordinate Bethe ansatz, already used in 
\cite{RIBEIRO} and we also leave the problem of counting of the spectral
multiplicities as an open question.

\section*{Acknowledgments}

This work was supported in part by Brazilian Research Council (CNPq), grant
\#310625/2013-0 and FAPESP, grant \#2011/18729-1.

\end{document}